# Reforming Physics Exams Using Openly Accessible Large Isomorphic Problem Banks created with the assistance of Generative AI: an Explorative Study

Zhongzhou Chen, Emily Frederick, Colleen Cui [b], Munaimah Khan, Christopher Klatt, Mercedith Huang, Shiyang Su [b]

[a] *Department of Physics, University of Central Florida, 4111 Libra Drive, Orlando, Florida, 32816*

[b] *Department of Psychology, University of Central Florida, 4111 Pictor Lane, Orlando, Florida, 32816*

This paper explores using large isomorphic problem banks to overcome many challenges of traditional exams in large STEM classes, especially the threat of content sharing websites and generative AI to the security of exam items. We first introduce an efficient procedure for creating large numbers of isomorphic physics problems, assisted by the large language model GPT-3 and several other open-source tools. We then propose that if exam items are randomly drawn from large enough problem banks, then giving students open access to problem banks prior to the exam will not dramatically impact students' performance on the exam or lead to wide-spread rote-memorization of solutions. We tested this hypothesis on two mid-term physics exams, comparing students' performance on problems drawn from open isomorphic problem banks to similar transfer problems that were not accessible to students prior to the exam. We found that on both exams, both open bank and transfer problems had the highest difficulty. The differences in percent correct were between 5% to 10%, which is comparable to the differences between different isomorphic versions of the same problem type. Item response theory analysis found that both types of problem have high discrimination (>1.5) with no significant differences. Student performance on open-bank and transfer problems are highly correlated with each other, and the correlations are stronger than average correlations between problems on the exam. Exploratory factor analysis also found that open-bank and transfer problems load on the same factor, and even formed their own factor on the second exam. Those observations all suggest that giving students open access to large isomorphic problem banks only had a small impact on students' performance on the exam but could have significant potential in reforming traditional classroom exams.

# I. INTRODUCTION

The rapid development in both the capability and accessibility of generative AI models and services such as BERT [1], GPT-3 [2], and ChatGPT (chat.openai.com) are poised to have profound impacts on many aspects of life and work. The ability of generative AI models to understand and generate human-like text enables a broad spectrum of applications from writing assistance to automated customer services and beyond. At the heart of the latest generative AI technologies are pre-trained large language models (LLMs): models that are trained on a large corpus of text data drawn from the internet and learn to predict the probability of the next word in a sentence given the preceding words[1]. Pre-trained LLM models are capable of generating text that is contextually relevant and coherent, given some initial input or 'prompt'. More importantly, they can achieve high performance in a wide variety of text generation tasks by learning from only a few examples (few shot learning) [3].

There is an increasing body of research on applying LLMs to solve a wide range of problems in education, ranging from automated or assisted scoring of student short answers [4–9], generating feedback and dialogue on student answers [10] to using conversational AI to support language learning [11]. In particular, LLMs have been applied to the automatic or assisted generation of problems used in both practice and assessment settings. For example, Dijkstra et. al. trained a GPT-3 model to generate multiple choice questions with fair answers and distractors [12]. Jiao et. al. utilized an "energy-based" model for automatic generation of math word problems of appropriate difficulty level [13]. Shimmei et. al. incorporated learning objectives into the problem generation process and generated questions evaluated to be on par with those generated by human experts in terms of quality [14]. Automatic or assisted problem generation has been shown to be more cost effective than manual writing when a large number of problems are created [15], and as LLMs become more widely available and less costly to implement the cost of problem generation will continue to decrease.

Given generative AI's potential to significantly lower the cost and effort of problem generation, a key question that we seek to answer in this paper is how to properly harness this power to transform and improve existing practice of teaching and learning. In this paper, we propose and evaluate one possible way to overcome the challenges facing conventional classroom exams in large introductory STEM classes, utilizing large banks of isomorphic assessment problems efficiently created with the assistance of LLM and other open source tools.

**Challenges of Conventional Exams in STEM Classes**

In most high-school and college introductory level STEM classes, each exam theoretically requires all students in the class to complete a single set of problems at the same location and time, due to two reasons:
1. To ensure fairness for every test taker by giving everyone the same set of problems.
2. To ensure that no test taker can have an unfair advantage by gaining access to the problems ahead of time.

As a result, students should presumably not have access to the exact same problems prior to the exam. In that case, the exam results would reflect students' ability to transfer their knowledge to new situations, rather than their ability to rote memorize the answer or the solution to a specific set of problem.

Despite being widely accepted as the standard practice for assessing student learning, this form of "secure" and synchronous exam is facing an increasing number of challenges, especially in larger classes. First, exams are increasingly becoming a significant accessibility hurdle for students [16–18] as the higher-education population quickly becomes more diverse in socio-economic status, age, demographics and cultural backgrounds [19]. The non-traditional student population faces more challenges in trying to keep up with a rigid exam schedule. In addition, packing hundreds of students in one lecture hall also poses a health-risk for both instructors and students [20,21], especially high-risk populations, by increasing the risk of exposure to contagious pathogens such as COVID and flu. Second, in large STEM gateway courses with hundreds of students, administering a synchronous exam can be logistically challenging for instructors, which prevents the administration of more frequent and lower stakes assessments in the course that can be beneficial for learning [22,23].

At the same time, the ability of instructors to uphold the security of assessment problems is being dramatically compromised by the proliferation of online resource sharing websites such as Chegg, CourseHero, or online discussion platforms such as Reddit and Stack Exchange [24–26], and most recently by the prevalence of AI tools such as ChatGPT [27,28]. This creates a serious dilemma for instructors and education researchers: On one hand, if the same set of assessment problems are being used in multiple classes and reused over a period of time, then some students can easily gain access to the problems prior to the exam, thus violating the secure items requirement of conventional exams. On the other hand, if a new set of problems are being created and used for each new assessment, it will be very challenging to keep the quality and difficulty of the problems relatively the same across exams, and make it difficult to compare student exam performance over time or between different classes. It will also be challenging to improve the quality of the problems based on students' assessment data since they are used only once.

---
[1] The LLM prediction is made based on tokens which loosely corresponds to 2-3 letters.

**Reforming Exams Using Large Open Isomorphic Problem Banks**

Since all the above-mentioned challenges arise from the "secure" requirement of traditional exam, one way to overcome those challenges is to remove this requirement by making exam problems openly accessible to all students prior to the exam. The challenge is that if the exam only involves a relatively small number of problems, then open access to exam problems will enable test takers to rote memorize the answers. Given generative AI's growing ability to create large number of problems, we propose the following simple method of both making assessment problems openly accessible while also discouraging rote-memorization of answers:
1. Each exam problem will be drawn from a large bank of isomorphic problems, created with the assistance of generative AI.
2. The large isomorphic problem banks are openly accessible to students prior to the exam to study and practice (but not accessible during the assessment).

The validity of this approach is based on two underlying assumptions:
1. When each problem bank is large enough, then rote-memorization of solution will be a much less efficient, almost impractical strategy, compared to understanding of the problem-solving process. Thus, large open isomorphic problem banks will discourage rote learning.
2. When the problems in the problem bank are similar enough to each other (i.e., isomorphic), then the exam will still be relatively fair, since every problem tests the same set of knowledge or skill, and the difficulty of each problem will be similar.

A core concept of this approach is "isomorphic problems". Many different researchers have proposed different definitions of isomorphic problems. For example, Hayes and Simons defined isomorphic problems as "problems whose solutions and moves can be placed in one-to-one relation with the solutions and moves of the given problem." [29] Millar and Manoharan loosely defined isomorphic problem to be "problems [that] are isomorphic for a targeted learning outcome and concept" [30]. In physics, multiple studies have considered two problems to be isomorphic if the problems "require the same physics principle to solve them." [31–34].

For our current purpose of creating large exam problem banks, the definition by Hayes and Simons is too restrictive since there needs to be a clear one-to-one mapping of the steps of the solution of the one problem to another [29]. This would likely promote rote memorization of solutions since students will quickly find out that all problems in a bank can be solved by using the same formula or solution procedure. On the other hand, the more common definition in physics education runs the risk of labeling problems with large differences in difficulty as "isomorphic". For example, Lin and Singh [35] considered problems involving two or three steps solutions as isomorphic, which are very likely to have big differences in difficulty for the average college introductory level students. Therefore, in the current study we chose our definition of isomorphic problems to be more generous than Simon and Hayes and more stringent than that of Lin and Singh.

We define isomorphic problems as problems that are being created from a common "seed" problem by applying a set of isomorphic variations. Each isomorphic variation should: 1) preserve the main concepts required to solve the problem, 2) preserve the overall complexity of the solution, such as the number of steps or the type of mathematical operation, and 3) introduce one or more minor changes to the solution that are less likely to affect its difficulty. For example, with a few exceptions, rotating the known and unknown variables in a problem can be considered an acceptable isomorphic variation for most college level students. Flipping the direction of an applied force from pushing to pulling will in most cases lead to a sign difference in the solution formula, which can also be considered isomorphic modification. A more rigorous and comprehensive definition of "isomorphic variation" is beyond the scope of the current study and will be further elaborated in the Discussion section. This definition is closest to Millar and Manoharan's definition that isomorphic problems involve the same targeted learning outcome and concept, with the addition that the overall complexity of the solution structure be preserved between problems.

**Research Questions and Hypothesis**

In this paper we seek to answer two main questions regarding the open isomorphic problem bank approach. First, is it feasible to efficiently create large enough isomorphic problem banks with the assistance of generative AI and other open-source tools? We will demonstrate the feasibility by explaining in detail in the Methods section our current process for creating large numbers of isomorphic problems assisted by GPT-3 and other free or widely accessible tools, and discuss possible methods of further improving the efficiency and quality of the process in the Discussion section.

Second, to what extent will making the isomorphic problem banks openly accessible to students promote rote learning, and artificially reduce the problem's difficulty when administered on an exam? In other words, will students' performance on those problems reflect their mastery of transferable knowledge, or merely their ability to memorize solutions? To answer this question, we will compare student performance on two types of problems on two exams administered in a traditional, synchronized fashion:

1. *Open bank problem:* each open bank problem is randomly drawn from one open isomorphic problem bank that was made accessible to students prior to the exam.
2. *Transfer problem:* each transfer problem is highly similar to and can sometimes be considered as isomorphic to the open bank problems, but students will not have access to the problems prior to the exam.

If having open access to a problem bank encourages students to rote learn or gain shallow, specific knowledge, then students will have much higher chances of success on the open bank problems on the exam than on the transfer problems. Moreover, students' performance on these two types of problems should be uncorrelated or weakly correlated, since students will solve the former using rote memorization, while having to solve the latter using their conceptual understanding.

More specifically, we hypothesize that if open bank problems would significantly increase rote learning among students and impede their mastery of transferable problem knowledge, we should observe the following on students' exam performance:

**Hypothesis 1**: The open bank problems will be significantly easier and have less discrimination power compared to the transfer problems. Furthermore, the difference in measured difficulty between the two types of problems will be much larger than the measured differences in difficulty within separate versions of either the open bank or the transfer problems.

**Hypothesis 2**: Students' performance on the open bank problems will not be correlated to their performance on the transfer problems, nor with their performance on other problems on the exam, because students are using a different method (most likely rote learning) to solve those problems.

**Hypothesis 3**: If both the open bank problem and the transfer problem are randomly chosen from two or more isomorphic versions on the exam, then students' performance on both problems should be more correlated when the two versions that they receive share more common surface features, and less correlated when they receive two versions with less common surface features.

H3 is based on previous research on the "specificity effect" [36]: when students' understanding of a concept is relatively shallow, their ability to transfer to a new context can be impacted by similarity of extraneous surface features of the two contexts. For example, students who learned to solve a problem in which a box moves to the right will be more successful in solving a problem in which an object also moves to the right, compared to when the object moves to the left, even though the solutions are considered isomorphic by experts. Therefore, if students only gained shallow understanding from studying the problem bank, then they will be more likely to transfer the knowledge to a problem with more similar surface features.

**Structure of the remainder of the paper**

The Methods section is divided into three parts. In part one, we introduce in detail our process of generative AI assisted isomorphic problem creation. In part two, we describe the instructional context in which the problem banks are being implemented, and the overarching design of the two studies. In the last part, we explain the analysis methods used to measure the difficulty and discrimination of problems, as well as the correlation between different problems and different versions of the problems.

The next two sections each report on one study. Each section starts with explaining the design and implementation details of the study, followed by the results of data analysis on the given study, and a brief summary of the key results from the study. A general discussion of the results and their implications is presented in the Discussion section, which concludes with limitations of current study and promising future directions.

## II. METHODS

**Generative AI Assisted Creation of Isomorphic Problem Banks**

The current process of creating an isomorphic problem bank of numerical problems consists of five steps as illustrated in the Figure 1. Steps 3, 4 and 5 are frequently conducted in parallel.

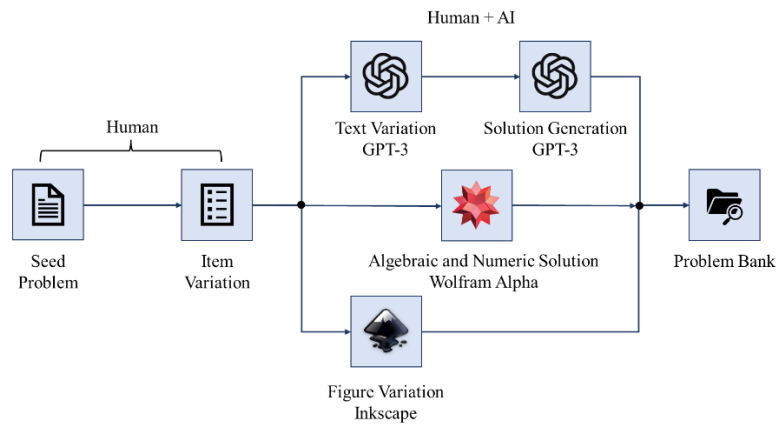

Figure 1: Schematic representation of the overall process of Generative AI assisted isomorphic problem generation.

*Step 1: Creating a "seed" problem.* A human expert first writes a "seed problem" that would serve as the basis of the isomorphic problem bank, which involves the learning objective(s) that the bank intends to assess. In practice, the seed problem is often a problem that the instructor would expect students to be able to solve towards the end of a chapter and could generate several isomorphic variations. The seed problem is then reviewed by more than one content expert to evaluate its clarity, appropriateness of difficulty, and adherence to the desired learning outcomes. A full problem text, problem diagram as well as a complete solution is created by the author at this step. A sample seed problem is shown in the first panel of Figure 2.

*Step 2: Identify Acceptable Variations.* The author(s) then determines a number of variations to the seed problem that could be seen as isomorphic within the scope of the course. In the current study, the isomorphic variations come in three hierarchical levels:
1. Major Variations: Change in problem context and/or minor sub-skills required.
2. Minor Variation: Change of minor details such as the orientation of a force or the direction of the motion of an object.
3. Rotation of Variables: Rotate the known and unknown variables in a problem.

For example, for the seed problem shown in Figure 2, a major variation is shown in the middle panel. In this variation, the context changed from a block to a cart, and the compression of a spring is replaced by the extension of a rubber band. In other cases, major variation could involve replacing sub-skills, such as replacing gravitational potential energy with elastic potential energy. An example of a minor variation is shown in the right most panel in Figure 2, in which the direction of motion is reversed from the seed problem. Minor variations typically correspond to small changes in the solution formula such as a change in positive and negative sign of a term.

Both major and minor variations are exclusively decided by human subject experts, who collectively determine if each variation can be conceived as isomorphic and will likely be of similar difficulty for the current student population and current instructional context.

*Step 3: Creation of problem figures assisted by Inkscape:* Creation and modification of problem diagram can be expedited by generating the original problem figure in scalar vector graphics (SVG) format. The SVG format allows individual elements of a figure, such a line, a box or a spring to be moved, copied, rotated and modified at any stage. For example, the image in the "Minor Variation" panel of Figure 2 can be simply obtained by moving the block and "compressing" the spring from the image created for the seed problem. SVG figures can be created and edited using a free open-source tool named Inkscape (https://inkscape.org/), and can be imported and edited in Microsoft PowerPoint. SVG figures are widely supported by all major web browsers and browser-based applications. In addition, SVG figures can be enlarged without pixelation, which can make the problem diagram more accessible for visually challenged students.

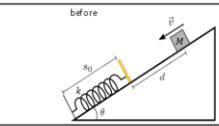

| | Seed Problem | Major Variation | Minor Variation |
|---|---|---|---|
| Image | (diagram: block on ramp, spring at bottom, labeled before) | (diagram: small cart on ramp with rubber band, labeled start) | (diagram: block on ramp, spring at bottom, labeled before) |
| Prompt | Block on ramp, spring at bottom. Given mass m, given theta, given spring length s_0, given speed v, given spring extends to s, given a distance d, find the spring constant k | Small cart on ramp tied to a rubber band, extend rubber band, cart mass m, theta, initial velocity of v_0 when the rubber band at initial length l_0, pulls rubber band to maximum l, Find spring constant of rubber band? | Block on ramp, spring at bottom. Given mass m kg, given theta, length s_0 with spring constant k, speed v, extends to s , a distance d. What is the length of the spring before the launch |
| Problem Text | A block of mass m = 4.70kg slides down from an inclined frictionless ramp that forms an angle of 36.19 with the ground. A long spring with a relaxed length s_0 = 0.65m is situated at the base of the ramp. The block started at a distance d = 2.11m from the tip of the relaxed spring, with an initial velocity v = 4.23m/s down the ramp. It compressed the spring to a length of s = 0.19m before temporarily coming to a stop. What is the spring constant of the spring? Round your answer to 2 decimal places. | You are designing a small cart and ramp system with a rubber band. The cart has a mass of m = 0.2kg, and it is tied to a rubber band. The ramp is at an angle of 25 degrees with the horizontal. After being released from the top of the ramp, the velocity of the cart reaches v_0 = 6 m/s when the rubber band is at its initial length of l_0 = 0.2m The rubber band is then pulled to a maximum length of l = 0.8 m. What is the spring constant of the rubber band? Round your answer to 2 decimal places. | A block of mass m = 12.35kg is resting against a spring at the bottom of a frictionless ramp that forms an angle of 10.98 degrees with the ground. The spring is initially compressed and has a length of s = 0.43m with a spring constant k = 463.10N/m. After the block is launched, the block travels a distance of d = 3.00m along the ramp from the tip of the relaxed spring, and its velocity is v = 5.13m/s before coming to a stop. What is the length of the spring after the launch? Round your answer to 2 decimal places |
| | (1/2)*k*(s-s_0)^2 = (1/2)*m*v^2 + m*g*(s_0-s+d)*sin(theta) | 1/2 k (l-l_0)^2 + m g (l - l_0) (sin(theta)) = 1/2 m v^2 + 0 | (1/2)*m*v^2 + m*g*(s_0-s+d)*sin(theta) = (1/2)*k*(s-s_0)^2 |

Figure 2: An example seed problem and relevant major and minor variations of seed problem. Each row represents one step in the creation process, starting with SVG image, human generated GPT-3 prompt text, GPT generated problem text, and Wolfram Alpha input for the three variations.

*Step 4: Generative AI Assisted Problem Text Writing:* Writing of isomorphic problem text is assisted by Generative Pre-trained Transformer 3 (GPT-3), an autoregressive language model with 175 billion parameters [28]. In the current study, access to GPT-3 is provided through the OpenAI website https://platform.openai.com/playground. To generate isomorphic variation of problem text, the model is set to "completion" mode. In this mode, the model is given a piece of text, and attempts to complete the text by predicting the most possible text that follows from the given text.

To generate the problem text for one isomorphic problem variation, the GPT model is first provided with a "prompt" for the seed problem followed by the seed problem text itself as an example. A prompt is a brief description of the essence of the problem, such as those listed in the "Prompt" row of Figure 2. A new prompt describing the first isomorphic variation is also included at the end of the input text. In other words, the initial input text has a "prompt-problem-prompt" structure. When submitted, GPT-3 will attempt to generate the first isomorphic variation problem text according to both the seed prompt-problem text pair, and the isomorphic variation prompt at the end. This procedure is referred to in related literature as one-shot learning [3].

The problem author then reviews the generated problem text and makes edits when necessary. The author then appends a new variation prompt after the previously generated text. GPT-3 will then generate the new variation problem text, using the two previous prompt-text pairs as examples (which is called few-shot learning). In practice, the problem author would start with rotation of variable variations, and generate 5-7 prompt-text pairs as example, then attempt to generate minor or major variations based on those examples. Problem solutions were also generated via the same process.

Note that there are no requirements for the format or length of the problem prompt and prompts that are much shorter than those in Figure 2 have been shown to work well in many cases. All the problem variations generated for the current study are completed using free credits provided by the OpenAI website upon registration. The estimated cost for generating all the problem text used in this study is less than $1.

*Step 5: Creating correct answer formulae or number assisted by Wolfram-Alpha:* To produce the correct numerical or symbolic answer to a problem, a content expert modifies the solution formula of the seed problem according to the problem text of each variation, and inputs the equation or equation set into Wolfram Alpha, an online computational knowledge engine developed by Wolfram Research, which performs the necessary algebraic and numeric computation to obtain either a symbolic expression or a numeric value for the unknown variable. Using Wolfram-alpha significantly reduces the workload as well as the error rate of a human author.

All created problems are stored in a JSON file format developed by the research group and uploaded to a Github repository.

**Implementation of Open Isomorphic Problem Banks in Introductory Physics Course**

**Instructional Condition**: Open isomorphic problem banks were used as both exam practice material and assessment problems in a calculus-based university introductory-level physics class in Spring 2023 . The class had 328 registered students, of which 26% were Female, 32% were under-represented minority in STEM, 17% were first generation students, and 21% were transfer students from 2-year institutions.

The course was taught in a blended instruction mode: students were instructed to view pre-recorded lecture videos and conduct online homework using the Obojobo Next online learning platform [37]. Class meeting time was reduced from 3 hours to 2 hours compared to the traditional face-to-face version of the course, and was reserved for discussion of problem-solving skills and other interactive activities.

**Isomorphic Problem Banks**: Three isomorphic problem banks were created during the semester, as listed in Table 1. The problem banks were made available to students as not-for-credit practice activities one week prior to the corresponding mid-term exam (see detailed explanation below). Only the first two banks (Static Friction and Energy Conservation) were used in the current study. Data from the student usage of the third bank was not included due to a human error that accidentally made one problem variation significantly more difficult than the others.

Table 1: Number of major, minor, and rotation of variable variations, and the total number of isomorphic problem variations in each of the three isomorphic problem banks.

| Problem Bank Name | Major Variations | Minor Variations | Rotation of Variables | Total Problems |
|---|---|---|---|---|
| Static Friction | 3 | 2 | 4 | 24 |
| Energy Conservation | 4 | 2 | 5 - 6 | 44 |
| Ballistic Pendulum | 3 | 3 | 4 - 6 | 45 |

The seed problems for the first two problem banks are shown in Figure 3 and Figure 4 below:

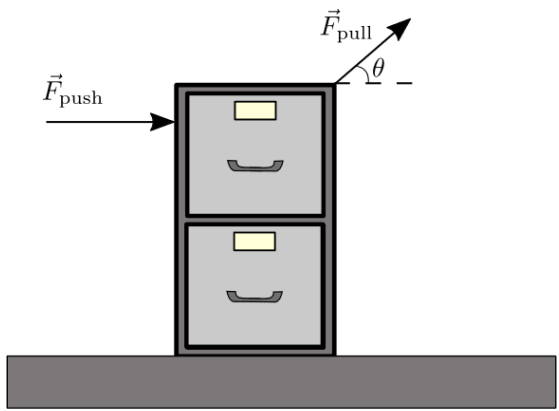

A friend is trying to pull a filing cabinet into their office. The cabinet has a mass of 70.0 $kg$ and they can't get it to budge on their own. You decide to help and begin pushing horizontally in the same direction. Your friend is pulling with a force of 200$N$ at an upward angle q = 20.0° from the horizontal. If the cabinet has coefficient of static friction $m_s$ = 0.620 with respect to the floor, what is the force you must exert to just get it moving? Retain your answer to two decimal places.

Figure 3: Human generated seed problem text and associated image for the static friction problem bank.

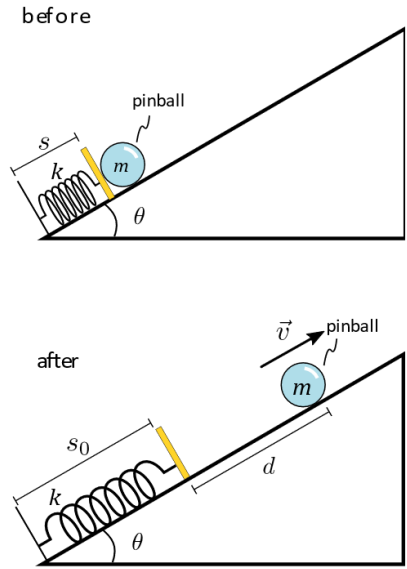

You are designing a rudimentary pinball machine and trying to find the perfect spring for your machine. The spring initially rests at $s_0 = 4.0\ m$. You want to use an antique pinball that you found that has a mass $m = 1.0.5\ kg$. Based on your design, you know that the angle of your playfield will be $q = 15.0°$, your pinball needs to have a final ejected velocity of $v = 1.8\ m/s$, and it needs to travel $d = 2.2\ m$ in order to enter the playing field properly. For simplicity, assume the pinball is not rolling and just sliding on the smooth field without friction. What does your spring constant need to be if you only have enough room for the spring to be compressed to a length of $s = 1.4\ m$? Round your answer to 2 decimal places.

Figure 4: Human generated seed problem text and associated image for the energy conservation problem bank.

The isomorphic problem banks were presented to students as online learning modules using the same Obojobo Next online learning platform used for online homework in the course, approximately one week prior to the exam date. Students have 1000 attempts on each module. For the Static Friction problem bank, 1 problem was randomly chosen for students on each practice attempt. For Energy Conservation, 3 problems were chosen on each practice attempt. The correct solution of the problem was revealed to students after each attempt.

**Administration of Exams**: A total of three mid-term exams were administered during the semester. The exams were administered synchronously during class times with an option to take the exam remotely with video camera on per student request. Each exam is conducted as an auto-graded Quiz on the Canvas Learning Management System [38]. All problems were either multiple-choice or numeric answer problems. All numeric answer problems had randomized variable numbers. Students were allowed 50 minutes to complete each exam. Approximately one week prior to the exams, the instructor made an announcement to the class on the topic of each problem to appear on the upcoming exam as a review guide. In Spring 2023 semester, the announcement explicitly pointed out that one problem would be directly drawn from one practice problem bank that students have access to, and another problem (two problems on exam 2) will be similar to the problems in the practice problem bank.

**Open-bank and Transfer Problems on Exams:** Two new problems were included on the mid-term exam 1: one open-bank problem, directly selected from the Static Friction problem bank, and one Transfer problem. For mid-term exam 2, one open-bank and two Transfer problems were included. Each problem was randomly drawn from two similar versions. A schematic representation of the study design including one open bank problem and one transfer problem is shown in Figure 5. To answer RQ3, the two transfer problems were designed so that version A of the transfer problem shared more surface features with version A of the open bank problem, such as having the same direction of motion, and vice versa. Details of the open-bank and transfer problems on each exam are presented in Study 1 and Study 2 sections respectively.

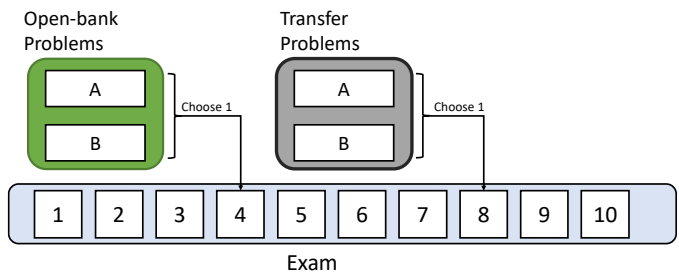

Figure 5: Schematic representation of the design of study 1 on Exam 1. Green box represent open-bank problems on problem 4, gray box represents transfer problems on problem 8. The design of study 2 follows a similar scheme

There were three reasons for only including two versions for each problem, rather than randomly selecting the open bank problem from the entire isomorphic problem bank:
1. The exam was administered synchronously, so students did not know which two problems would be selected from the bank.
2. Selecting two problems ensures sufficient amount of data on each problem for reliable measurement of correct percentage and proper fit of IRT model. It also greatly simplifies transfer analysis between open bank and transfer problems on RQ3.
3. Since the open problem banks are newly created and do not have student practice data, selecting two problems reduces the possibility of accidentally including one problem that accidentally has greater difficulty than others.

For each open bank problem and transfer problem, 10 sets of random numbers were generated for the variables in the problem to reduce the chances of direct answer copying between students.

**Data Analysis**

To examine hypothesis 1, difficulty of problems on the exams are measured by two methods: Classical Test Theory (CTT) and Item Response Theory (IRT)[2]. CTT is the traditional approach to evaluating scores on tests, where an examinee's ability is revealed through the total proportion of problems that they answered correctly [39]. In CTT, item difficulty is simply measured by the proportion of correct responses to a problem. The higher the proportion correct, the easier it is assumed to be. The statistical significance of differences between CTT difficulties of items can be tested with either a McNemar test (2 items) or a Friedman's test (more than 2 items), on the proportion of correct answers on each item [40]. CTT provides a straightforward estimate of each item's difficulty.

IRT is a set of models that predict latent traits (or ability level) using the probability of endorsing correct responses on assessment items. IRT is widely used nowadays in most large-scale, high stakes assessments [41], as well as analysis of students' performance in online courses and standardized research instruments [42,43]. IRT assigns an ability level ($\theta$) to each test taker on a latent continuum based on their exam performance. Theoretically, the latent continuum ranges from $-\infty$ to $+\infty$, while typical values of $\theta$ ranges from –3 to +3.

In IRT, the item difficulty (also called item location or the $b$ parameter), indicates the test taker's level of ability at which there is a 50% chance of answering an item correctly. This parameter is included in all IRT models. Difficult items are located to the right (or higher end) of the ability scale, while easier items are located to the left (or lower end) of the ability scale. Typical values of item difficulty $b$ range from −3 to +3. The interpretation of item difficulty IRT is the opposite of the item difficulty statistic in CTT—a lower value here indicates an easy item, and a higher value indicates a difficult item.

IRT models with more than 1 parameter can also estimate the discrimination of each item. Item discrimination (or the $a$ parameter) is defined as the measure of the differential capability of an item. It refers to how well an item can differentiate between examinees with different ability levels along the continuum of the ability scale. High item discrimination suggests that the item can easily differentiate between examinees with a high ability level vs. a low ability level.

In both studies, we fitted the exam data with 1, 2 and 3 parameter IRT models using the ltm R package [44] and compared fit indices. We conducted ANOVA tests and compared the log likelihood value, the Akaike Information Criterion (AIC [45]), and the Bayesian Information Criterion (BIC [46]) to determine the best fit model. The best fit model should have a higher log likelihood value, a lower AIC, and/or a lower BIC.

For hypothesis 2 and 3, correlation between open bank problems with transfer problems and other problems on the exam can be measured by the Pearson correlation coefficient and visualized using heatmaps. In addition, correlations between exam problems can be evaluated using Exploratory Factor Analysis (EFA). EFA is a technique used to identify the underlying dimensionality of data. It identifies the number of latent variables (factors) that explain the observed variables through extracting the maximum common variance among them. Each item will correlate to a certain extent with each factor, and this correlation is referred to as the factor loading. Items are considered to load primarily onto one factor when the value of the loading is above 0.3 and when it does not considerably load onto another factor. All items on the exam that load onto the same factor theoretically measure the same underlying knowledge or skill. In physics education research, EFA has been used to detect underlying knowledge structures in standardized assessments such as the force concept inventory [47].

---

[2] in the field of measurement/psychometrics, researchers refer to problems on an exam or questions on a survey as "items"

In this study, we performed EFA by using principal axis factoring and an oblimin rotation of the items. To determine the best number of factors to retain, we considered both the number of factors with eigenvalues over 1.00 and the inflection point in the scree plot [48].

In both studies, the open-bank problem and transfer problem are randomly drawn from two isomorphic versions, A and B. Therefore, each of the above analysis is conducted twice for each study: The first time ignoring the differences between the two versions and treating both versions as the same problem, the second time treating the two versions as separate problems. The only exception is the EFA analysis, which could not handle missing values. Therefore, it is only performed by treating the two versions as the same problem.

# STUDY 1

## Study 1 Setup

The first study was implemented in mid-term exam 1 which contains 10 problems covering topics ranging from vector algebra, projectile motion, to forces and Newton's laws of motion. Problem 4 on the exam was the open bank problem, for which each student was randomly assigned one of two problems (labeled Q4A and Q4B) [Figure 6]. Problem 8 was chosen to be the "Transfer" problem, and each student was randomly assigned one of two transfer problems, labeled Q8_A and Q8_B, shown in Figure 7.

Transfer problem Q8 is slightly more complicated than open bank problem Q4, in that Q4 has only one force that is at an angle to the horizontal whereas in Q8 both forces are at an angle to the horizontal. Transfer problem Q8_A was designed to be more similar to Q4_A in that both forces in both problems are directed outwards from the object, i.e. being pulling forces. In addition, in both problems forces at an angle to the horizontal direction are pointing upwards instead of downwards. Similarly, Q8_B is more similar to Q4_B than to Q4_A.

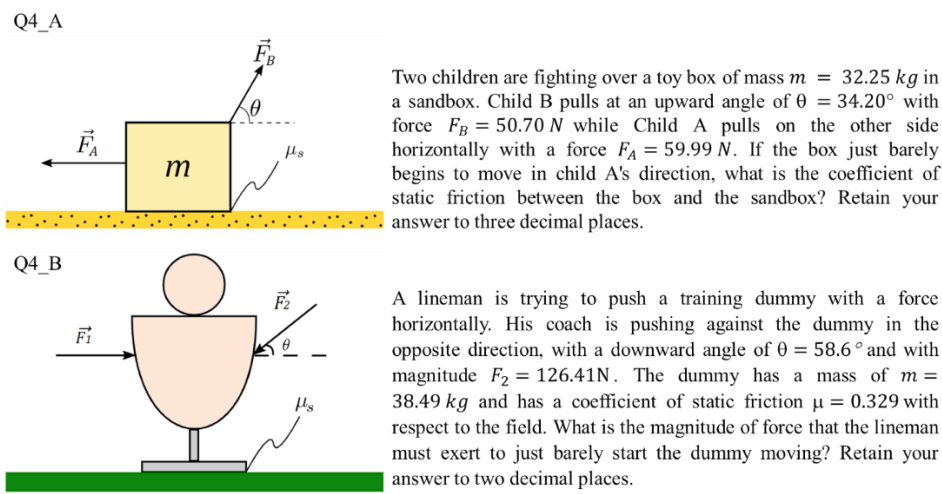

Figure 6: The two versions (Q4_A and Q4_B) of the open-bank problem (Question 4) on the mid-term exam 1. Both versions were selected from the static friction problem bank which was accessible to students as a study resource for a week prior to mid-term exam 1.

## Study 1 Results

A total of 292 students took mid-term exam 1
. The average correct percentage on all problems was 55%. A total of 37 students accessed the open problem bank prior to the exam.

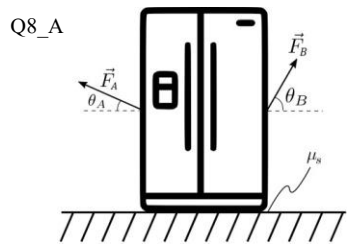

Q8_A

You and your roommate cannot agree on where to put a small to medium size fridge in your apartment, so the two of you started to tug the Fridge in two different directions. Your friend pulls the fridge with a force $F_B = 66.17\ N$, at an angle of $\theta = 56.54°$ above the horizon, but it is not enough to move the fridge. You pull the fridge with a force $F_A = 67.71\ N$, at an angle of $\theta = 14.47°$ above the horizon, which is just big enough that the fridge barely starts to move towards you. If the coefficient of static friction between the floor and the fridge is $\mu_s = 0.05$, what must be the mass of the fridge in units of $kg$? retain your answers to 2 decimal places.

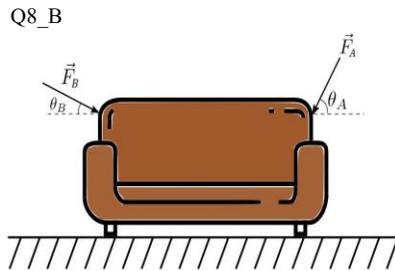

Q8_B

You and your roommate decide to move the sofa in your living room. You push the sofa with force $F_A = 47.84$ N, at an angle of $\theta = 59.73°$ above the horizon. However, due to miscommunication, your friend pushes at the same time, with force $F_B = 73.6\ N$ from the other side of the sofa, and an angle $\theta = 33.92°$ above the horizon. Your friend's pushing force is just big enough so that the sofa just barely starts to move towards you (in your friend's direction of pushing). If the coefficient of static friction between the sofa and the ground is $\mu_s = 0.07$, what is the mass of the sofa in units of $kg$? Retain your answer to two decimal places.

Figure 7: The two versions (Q8_A and Q8_B) of the Transfer problem (Question 8) on exam 1, which were randomly assigned to students using Canvas. These two problems were newly created as transfer problems to be comparable to questions Q4_A and Q4_B.

**Results from CTT:**

Figure 8 shows the proportion answered correctly for Q4, Q8, alongside the average proportion for all other items. Data from both versions A and B of Q4 and Q8 are combined in this plot. The error bars for Q4 and Q8 are binomial errors, whereas the error bars for all other problems on the exam are the combined binomial errors of those problems.

From Figure 8A, it is clear that Q4 and Q8 are much more difficult than the average problem on the exam, by 15% or more. Q8 (34%) also has a 10% lower correct rate than Q4 (44%), which is statistically significant (McNemar test, $\chi^2=10.26$, p<0.01, df=1).

Figure 8B shows the proportions answered correctly for both versions (A and B) of both Q4 and Q8. For both Q4 and Q8, version B is about 10% harder than version A. Q4_A has the highest correct rate of 0.38, whereas Q8_B has the lowest correct rate of 0.29. The difference between Q4_A and Q8_B is statistically significant (McNemar test, $\chi^2=9.33$, p=0.01, df=1, p-values adjusted using the Hochberg method). On the other hand, the correct rate on Q4_B and Q8_A are almost identical.

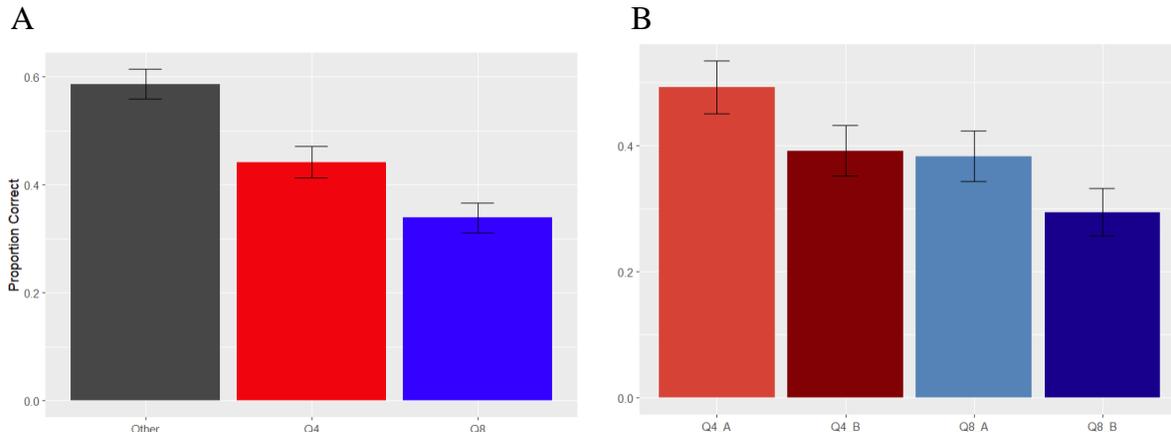

Figure 8: Proportion of correct answers for Q4 and Q8 (A) Comparison of the two test questions Q4 and Q8 to the average proportion of correct for the rest of the exam questions (other). (B) Comparison of the proportion correct for individual versions of Q4 (Q4_A, Q4_B) and Q8 (Q8_A, Q8_B).

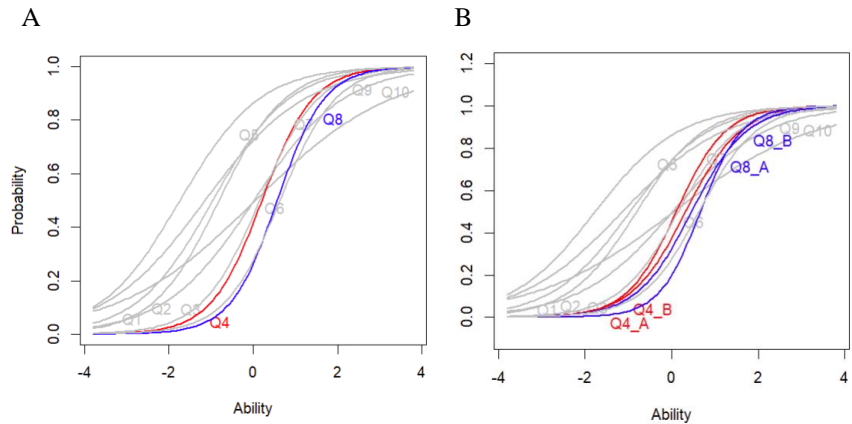

Figure 9: Item Characteristic Curves for mid-term exam 1. (A) ICC with Q4 and Q8 each considered as one item. (B) ICC with the two versions of Q4 (Q4_A and Q4_B) and Q8 (Q8_A and Q8_B) considered as different items.

**IRT Analysis:**

When data from both versions of Q4 and Q8 are combined, a 2-PL model is deemed the best fit for the data. The item characteristic curves of all problems are shown in Figure 9.

As seen in Figure 9, the IRT difficulty of Q4 (0.21) and Q8 (0.57) (center of item characteristic curves) are among the highest of all problems on the exam. The discriminations of both items (the slope of item characteristic curves) also tend to be similar or steeper than other items on the exam.

While the IRT difficulties between the two problems are significantly different (t-test, $t = -2.21$, $p = 0.03$), the difference is rather small compared to the variation of difficulties between those two and all other items on the exam, which range from 0.63 to -1.74. The discrimination of Q4 (1.65) is nearly identical to the discrimination of Q8 (1.80).

When the two versions of each problem are treated as separate items, a 2PL model is still the best fit for the data. In Figure 9B, we see that the difference in difficulties between separate versions of Q4 and Q8 are small compared to the differences in difficulties between those four items (Q4_A, Q4_B, Q6_A, Q6_B) and all other items on the same exam. There are no significant differences in discrimination between all problems on the same exam.

Table 2: Factor loading of the EFA analysis for the two-factor model of exam 1.

|     | Factor 1 | Factor 2 |
| --- | --- | --- |
| Q1  | **0.28** |       |
| Q2  | **0.54** | -0.14 |
| Q3  | **0.48** |       |
| Q4  | **0.52** |       |
| Q5  | 0.21     | 0.19  |
| Q6  | **0.37** | 0.24  |
| Q7  | -0.01    | **0.57** |
| Q8  | **0.55** |       |
| Q9  | 0.23     | **0.36** |
| Q10 | **0.50** |       |

**Correlation between problems**

Correlation coefficients between each pair of problems on the exam are shown in the heatmap in Figure 10A, which treats the two versions of Q4 and Q8 as single problems, and Figure 10B, which treats the two versions as separate problems. In Figure 10, Q4 and Q8 very clearly have the highest correlation among all problem pairs on the exam ($r = .43$). In Figure 10B, the highest correlation among all problems is between Q4_B and Q8_A ($r = .61$), and the second highest correlation is between Q4_A and Q8_B ($r = .47$), all of which can be considered as moderate to strong correlation [49]. The correlation coefficients between the mismatched versions on Q4 and Q8 are much larger than between the matched versions (Q4_A and Q8_A, Q4_B and Q8_B).

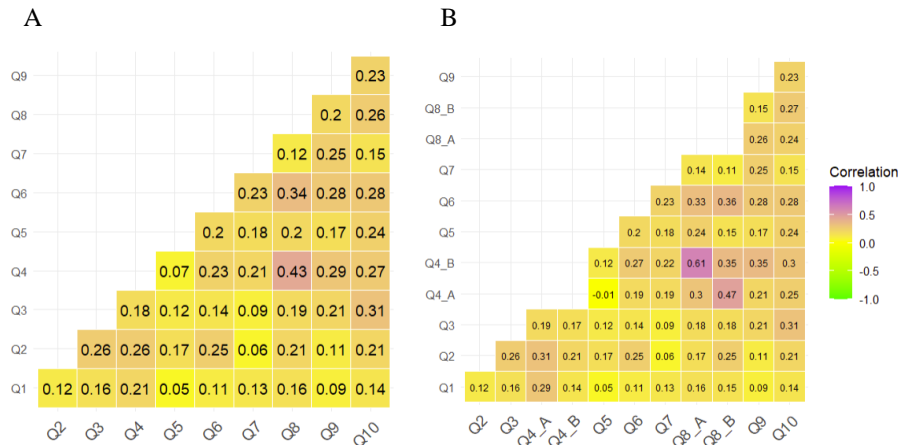

Figure 10: Heat map representation of correlation coefficients between each pair of problems on Exam 1.

**Exploratory Factor Analysis:**

The scree plot (Figure 11) shows that the exam data is best fitted with a two-factor model. As shown in Table 2, both Q4 and Q8 loads exclusively on factor 1, together with most of the other problems on the exam. The only problem that loads exclusively on factor 2 is problem 7, which is a conceptual question regarding projectile motion.

**Summary of Study 1 Results**

The results of Study 1 are summarized below, in terms of their relation to the three hypotheses we proposed in this study.

Hypothesis 1: We found mixed evidence regarding hypothesis 1, which predicted that the open-bank problem would be much easier than the transfer problem. On one hand, both CTT and IRT results confirmed that the transfer question Q8 is indeed more difficult than the open bank problem Q4. On the other hand, the difference in correct rate is only 10%, which is quite small compared to the difference in difficulty between all other problems on the same exam. Moreover, the differences in CTT difficulty between the two versions of both Q4 and Q8 are also about 10%, comparable to the difference between the two problems. Therefore, it cannot be concluded that the observed difficulty difference between Q4 and Q8 is the result of giving students open access to the isomorphic problem bank. Moreover, the discrimination values of both problems are nearly identical to each other.

Hypothesis 2: Contrary to hypothesis 2, which suggests that the open-bank and transfer problems will have weak correlations, we observed that that Q4 and Q8 are not only highly correlated, but also more correlated with each other than with any other problem on the exam. Exploratory factor analysis results also rejected the hypothesis by showing that Q4 and Q8 load on the same factor in a two-factor model.

Hypothesis 3: Hypothesis 3 suggests that matched versions of both problems will be correlated more with mismatched versions. On the contrary, we observed that mismatched versions of Q4 and Q8 correlated much more than matched versions. While the implications of this observation is unclear, it certainly didn't support the specificity hypothesis of student learning. However, the data did show that Q4_A and Q8_A, both which involve forces with positive vertical components, have higher correct rates than Q4_B and Q8_BT, which involve forces with negative vertical components. This might indicate that students are more likely to make a mistake under certain circumstances.

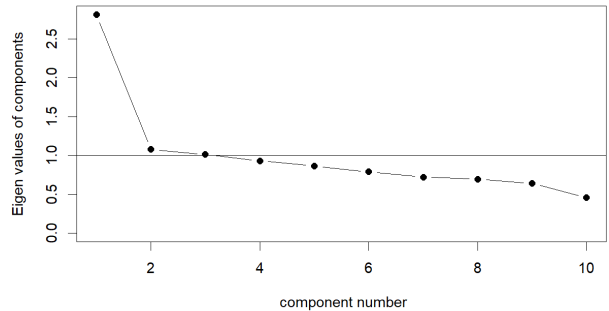

Figure 11: Scree plot of exploratory factor analysis on all problems on Exam 1.

Finally, there are several caveats to Study 1 that are worth mentioning. First, only a small number of students (37) accessed the problem bank for practice prior to the exam. One likely reason is that this is the first time such an open access format is introduced, and many students might not have fully understood that the exam same problems will be drawn from the problem bank. Second, the problem bank used in the current study only contained 23 problem variations. Also, each practice attempt only gave students one problem from the bank, which reduced the number of isomorphic variations each student would access. Third, the transfer problem is mathematically slightly more complicated than the open bank problem, which confounds the interpretation of the cause of Q4 being somewhat easier than Q8. Those caveats are addressed in study 2.

## STUDY 2

**Study 2 Setup**

Study 2 was implemented in mid-term exam 2, which contained 9 problems, covering the topics of circular motion, Newton's laws of motion, work and mechanical energy, conservation of mechanical energy, and conservation of linear momentum. More specifically, problems 2, 3, 4, 6 and 8 all related to the topic of work and mechanical energy. Question 6 was chosen to host the open-bank problem, randomly drawn from two isomorphic problems, labeled Q6_A and Q6_B, from the problem bank (Figure 12A and B).

Figure 12: The two versions (Q6_A and Q6_B) of the open-bank problem (Question 6) on the mid-term exam 2. Both versions were selected from the energy conservation problem bank which was accessible to students as a study resource for a week prior to the mid-term exam 2.

Question 4 contained the first transfer problem, randomly drawn from two versions, labeled Q4_A and Q4_B. Q4 is computationally less complicated than Q6 because students do not need to take the component of the distance traveled by the object to calculate the change in gravitational potential energy.

### Q4_A

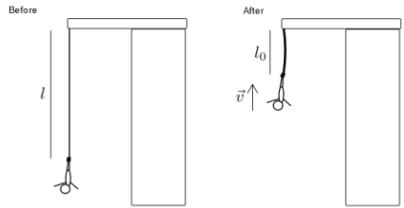

A bungee jumper that weighs $m = 63.04\ kg$ is jumping with an elastic bungee cord attached to his feet. After jumping off, the bungee cord extends to a length of $l = 26.3\ m$ when he temporarily stops and starts to bounce back. When the cord retracts to its relaxed length of $l_0 = 12.08\ m$, the jumper has an upward velocity of $v = 14.94\ m/s$. Assuming the bungee jumper bounces back in a straight line, what must be the spring constant of the bungee cord if we can model the cord as an ideal spring? Neglect air resistance and other types of friction and retain your answer to 2 decimal places.

### Q4_B

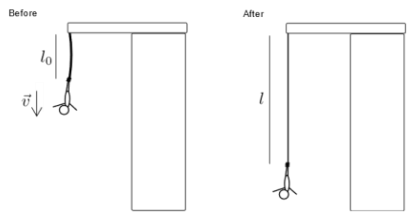

A bungee jumper that weighs $m = 84.72 kg$ is jumping with an elastic bungee cord attached to his feet. When the bungee cord first reaches its relaxed length of $l_0 = 10.17 m$, the jumper's downward velocity is $v = 22.66\ m/s$. When the bungee cord extends to a maximum length of $l = 21.69\ m$, the jumper temporarily stops and then starts to bounce back. What must be the spring constant of the bungee cord if we can model the cord as an ideal spring? Neglect air resistance and other types of friction and retain your answer to 2 decimal places.

Figure 13: The two variations (Q4_A and Q4_B) of Question 4 on the mid-term exam 2, which were randomly assigned to students using Canvas. These two problems were created as transfer problems to be comparable and slightly easier to questions Q6_A and Q6_B on mid-term exam 2.

Question 8 was the second transfer problem, randomly drawn from two problems, labeled Q8_A and Q8_B. Q8 is clearly more complicated than both Q4 and Q6 in that the mechanical energy is not conserved and require students to calculate the loss in mechanical energy due to work done by friction.

In comparison, Q4 can be seen as the near transfer problem to Q6, and Q8 as the far transfer problem to Q6.

### Q8_A

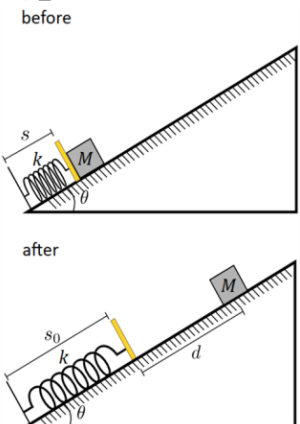

A block of mass $m = 2.86\ kg$ is resting against a spring at the bottom of a rough ramp that forms an angle $\theta = 20.86°$ with the ground. The spring is compressed to a length of $s = 0.62 m$. The spring has a spring coefficient of $k = 994.85 N/m$. When the spring is released, the spring extends to its rest length of $s_0 = 1.28\ m$ and the block is launched and travels a distance $d = 1.58\ m$ when it comes to a stop from the friction between the block and the ramp. Find the MAGNITUDE of work done by friction on the block in units of Joules (Do NOT include a negative sign in your answer). Round your answer to 2 decimal places.

### Q8_B

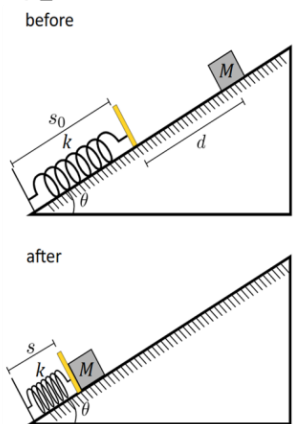

A block of mass $m = 5.44\ kg$ slides down from an inclined ramp that forms an angle $\theta = 26.95°$ with the ground. A long spring with a spring constant $k = 130.01\ N/m$ and a relaxed length $s_0 = 0.91\ m$ is situated at the base of the ramp. The block is released at rest from a distance $d = 2.31\ m$ from the tip of the relaxed spring. It compressed the spring to a length of $s = 0.52\ m$ when it first comes to a stop. Find the MAGNITUDE of work done by friction on the block in units of Joules (Do NOT include a negative sign in your answer). Round your answer to 2 decimal places.

Figure 14: The two variations (Q8_A and Q8_B) of Question 8 on the mid-term exam 2, which were randomly assigned to students using Canvas. These two problems were created as transfer problems to be a variation which is comparable to questions Q6_A and Q6_B on mid-term exam 2.

For all three problems, version A involves the "launching" process for which initial elastic potential energy stored in the system is converted into kinetic energy and gravitational potential energy. Version B involves the "capture" process for which initial kinetic and gravitational potential energy of the system is converted into elastic potential energy. However, the correct mathematical expressions for the two versions happen to be equivalent due to the scalar nature of mechanical energy.

Study II is designed to address the caveats of Study 1. First, students in the class had prior experience with the open bank format, therefore are more likely to access the problem bank for practice. The instructor of the class also made multiple announcements on one exam problem being directly selected from the problem bank. Second, Q4 is designed to be an easier transfer problem compared to Q6. As a result, if the observed difficulty of Q6 is significantly lower than that of Q4, it can be more straightforwardly attributed to the open problem bank format of Q6. Q8 can provide an upper limit for the estimated actual difficulty of Q6, which helps estimate the maximum change in difficulty. Finally, the problem bank is almost twice as big, which reduces the possibility of rote memorization or specific learning. Each student was also presented with two problems on each practice attempt, with each problem drawn from a different major isomorphic variation.

**Study 2 Results**

A total of 307 students took the second mid-term exam, with an average score of 52%. For the open problem bank, 175 students made at least one attempt on the practice module prior to the exam.

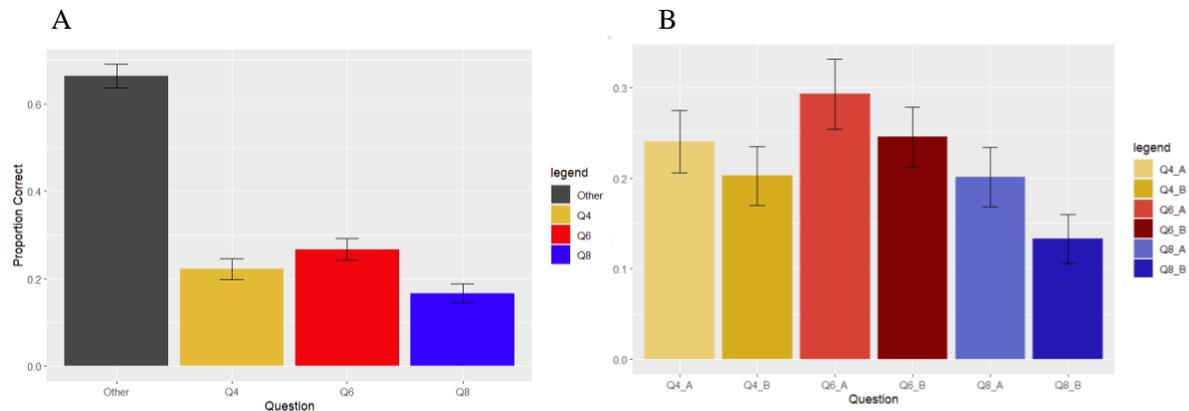

Figure 15: The proportion correct answers for Q4, Q6 and Q8 (A) Comparison of the test questions Q4, Q6 and Q8 to the average proportion f correct for the rest of the exam questions (other). (B) Comparison of the proportion correct for individual versions of Q4 (Q4_A, Q4_B), Q6 (Q6_A, Q6_B) and Q8 (Q8_A, Q8_B).

**Results from CTT**

Figure 15 shows the proportion of correct answers for Q4, Q6, and Q8 (with both versions combined), alongside the proportion correct for all other items. Q4, Q6, and Q8 are significantly more difficult than all other questions on the exam, as shown by the much lower rate of correct answers. A Friedman's test conducted on Q4, Q6 and Q8 showed that the correct rates of the three problems are significantly different from each other ($\chi^2 = 13.51$, $p < 0.01$, $df = 2$). Follow up pairwise McNemar tests showed that Q8 (17%) is significantly more difficult than Q6 (27%) ($\chi^2 = 15.25$, $p < 0.01$, $df = 1$), whereas the difference between Q6 (27%) and Q4 (22%) is not statistically significant ($\chi^2 = 2.01$, p = .16, df = 1)

Figure 15B shows the proportion correct for both versions (A and B) of Q4, Q6, and Q8. For each question, version B is slightly more difficult than version A, but none of the differences are statistically significant. Pairwise McNemar tests showed that Q8_B, the most difficult version, is significantly more difficult than Q6_A, the least difficult question ($\chi^2 = 8.45$, $p = 0.01$, $df = 1$), with p-values adjusted using the Hochberg method (Chen et al., 2017).

Table 3: This table shows the difficulty and discrimination parameters across the questions on the mid-term exam 2.

|    | Difficulty | Discrimination |
|----|------------|----------------|
| Q1 | -1.96      | 0.47           |
| Q2 | -0.93      | 1.52           |
| Q3 | -0.62      | 1.39           |
| Q4 | 1.27       | 1.28           |
| Q5 | -0.73      | 0.80           |
| Q6 | 0.94       | 1.51           |
| Q7 | -0.40      | 1.00           |
| Q8 | 1.16       | 2.62           |
| Q9 | -0.91      | 0.97           |

**IRT Analysis:**

The 2PL IRT model is shown to be the best fit for the data. The item characteristic curves (ICCs) for all questions, with Versions A and B combined, are shown in Figure 16.

The difficulty parameters of Q4, Q6, and Q8 in the combined test (Table 3) are each significantly higher than the difficulties of all the other items ($p < .01$). The discrimination parameters of the three target items, however, are more similar to those of the other items, with Q1 (0.47) being the only item with significantly lower discrimination than any target items—Q1 is significantly lower in discrimination than Q4 ($p = .01$) and Q8 ($p = .01$), calculated with a Hochberg p-value adjustment.

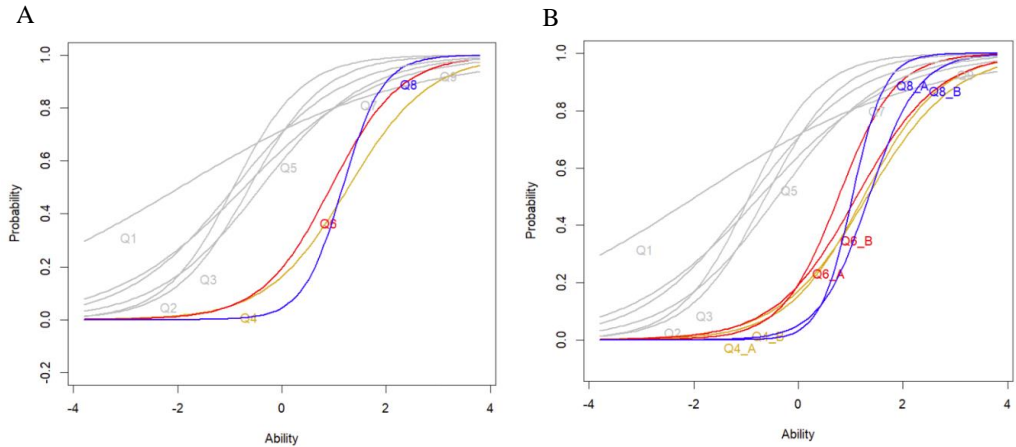

Figure 16: Item Characteristic Curves for mid-term exam 1. (A) ICC with Q4, Q6 and Q8 each considered as one item. (B) ICC with the two versions of Q4 (Q4_A and Q4_B), Q6 (Q6_A and Q6_B) and Q8 (Q8_A and Q8_B) considered as different items.

When the two versions of each question are treated as separate items, the 2PL model is still the best fit for the data. From the ICCs shown in Figure 16B, it can be seen that the difficulty parameters of both versions of Q4 (1.25 and 1.32), Q6 (0.77 and 1.12), and Q8 (1.03 and 1.35) are each still much higher than those of the other questions on the exam ($p < .01$). On the other hand, none of the discrimination parameters are significantly different between the three target questions and the other questions.

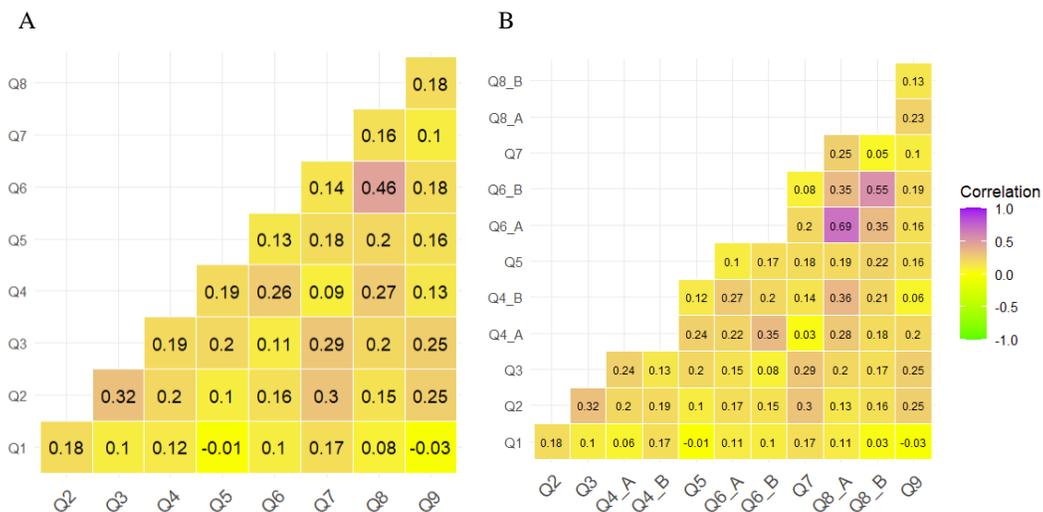

Figure 17: Heat map representation of correlation coefficients between each pair of problems on Exam 2.

Table 4: Factor loading of the EFA analysis for the two-factor model of Exam 2.

|     | Factor |      |
| --- | ------ | ---- |
|     | 1      | 2    |
| Q1  | 0.21   |      |
| Q2  | **0.59** |    |
| Q3  | **0.60** |    |
| Q4  | 0.17   | **0.33** |
| Q5  | 0.21   | 0.18 |
| Q6  |        | **0.67** |
| Q7  | **0.48** |    |
| Q8  |        | **0.69** |
| Q9  | 0.28   | 0.15 |

**Correlation between Problems**

Correlation coefficients between each pair of problems on the exam are shown in Figure 17A, which treats versions A and B as the same problem, and Figure 17B, which treats the two versions as separate problems. Of all the other problems on the exam, the open bank problem Q6 has the highest correlation coefficient with the near transfer problem Q4 (r = .26) and the far transfer problem Q8 (r = .46). The correlation between Q6 and Q8 is much higher than all other correlations between different problems on the same exam. The correlation between Q4 and Q6 (r = .26) is one of seven pairs of problems that have correlation coefficients of $r > .25$, whereas all the other problems had correlation coefficients of $r < .2$. Notably, the correlation coefficient between Q4 and Q8 is $r = .27$, which is essentially identical to the correlation coefficient between Q4 and Q6.

In Figure 17B it is shown that matching versions of Q6 and Q8 have higher correlations compared to miss-matched versions. On the other hand, for Q6 and Q4, mismatched versions have higher correlations than matched versions.

**Exploratory Factor Analysis:**

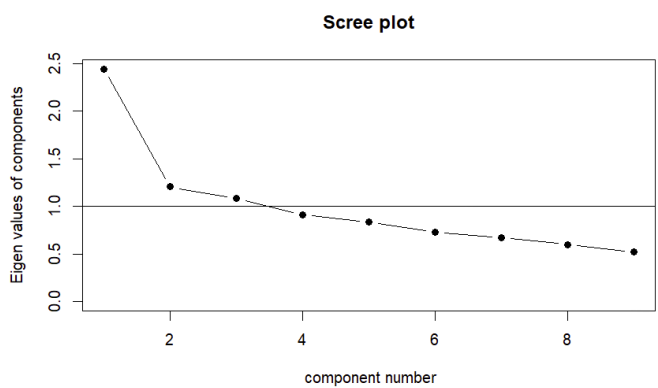

Figure 18: Scree plot of exploratory factor analysis on all problems on Exam 2.

Scree plot from the EFA analysis concluded that that a two-factor model is the best fit for the data (Figure 18). Factor loadings of the model is shown in Table 4 with Q4, Q6, and Q8 loading exclusively on Factor 2, and most other problems loads exclusively on Factor 1. Q5 and Q9 seems to load evenly on both factors, slightly favoring factor 1. The results are consistent with the observed large difficulty differences between the Q4, Q6, Q8 and all other problems on the exam.

**Summary of Study 2 Results**

A total of 163 students accessed the Energy Conservation open problem bank prior to exam 2, which is significantly more compared to exam 1. Yet the outcomes from Study 2 are similar to what was observed for Study 1. We again summarize the results of Study 2 according to the three hypotheses proposed in the introduction section.

Hypothesis 1: The results from CTT and IRT analysis on Study 2 largely rejects Hypothesis 1, since all three problems, Q4, Q6 and Q8, are much harder than all other problems on the exam. While Q8 is 10% more difficult than Q6, the difference was expected since Q8 requires more concepts to solve and the calculation is more complicated. On the other hand, the 5% difference between Q4 and Q6 was not statistically significant. Since Q4 was designed to be easier than Q6, and Q8 was designed to be hard, the results set the boundary on the estimated impact of the open problem bank approach to be between 5% and 10%. Note that this variation is comparable to the observed variance in difficulty between different isomorphic versions of the same item.

Hypothesis 2: is also rejected by the results of correlation and EFA analysis. Correlation between Q6 and either Q4 and Q8 are among the highest correlations between all pairs of problems on the exam. Result from EFA clearly shows that the three problems load on one factor, while all other problems load on a separate factor, suggesting that students are using similar procedures to solve all three problems, rather than using different approaches.

Hypothesis 3: Evidence for Hypothesis 3 is mixed in the results of Study 2. On the near transfer pair (Q4 and Q6), the mismatched versions are slightly more correlated than the matched versions, similar to what was observed in Study 1. On the far transfer pair (Q6 and Q8), the correlations between matched versions are stronger than mismatched versions. It is worth pointing out that even though Q6 is more conceptually aligned with Q4 than with Q8, Q6 and Q8 have more shared surface feature. In particular, the problem figures of Q6 and Q8 share a lot of common visual elements such as ramp and spring. It is well documented that novice students are more likely to categorize problems based on surface features [50,51]. Therefore, it is possible that the similarity of surface features between the two problems prompted many students to use largely identical procedures to solve both problems, leading to the significantly higher correlation between matched versions. The increase in complexity of Q8 might have also caused more students to be misled by surface features due to a high demand of cognitive resources. Finally, the fact that Q8 was after Q6 on the exam made possible for students' solution on Q8 to be "primed" by Q6.

## DISCUSSION

In this paper, we first demonstrated that it is feasible to generate isomorphic problem banks for college introductory level physics with around 30 to 50 acceptable variations. The utilization of Generative AI and SVG figure format improved the efficiency of problem writing to make the process practical.

Compared to other reported attempts of using generative AI to automatically generate large number of assessment items, the procedure that we developed requires significantly more human input at every step of the creation. The benefit of heavy human input is two-fold. First, it significantly reduces the chances of large language models generating erroneous content via "hallucination" [52–54], which can happen particularly frequently in natural science domains such as physics. Second, it could enable human creators to intentionally avoid and correct for possible biases that are inherent in LLM models trained on internet data [55–57].

While the effort required to create one problem bank is still significantly higher than creating a single problem, we argue that the additional workload can be more than justified by two unique benefits of the open nature of the isomorphic problem banks. First, open problem banks can be collaboratively created and shared among a large number of instructors, which reduces the workload on each individual. Second, since the banks do not need to be frequently renewed to prevent breaching of item security, they can be reused over many years, which reduces per-usage cost of creation.

It is worth pointing out that there is still much room to further improve the efficiency beyond the current process. For example, it is possible to ask GPT to generate all possible "rotation of variable" variations at once, and create all the acceptable variations with fewer submissions, by using more sophisticated prompt texts (which is known as "prompt engineering"). In addition, with the recent integration of Wolfram Alpha and GPT, the problem text, solution and explanation could be generated at once following a single prompt. Moreover, it is possible that future generative AI models could generate or modify problem diagrams following textual prompts. In short, the current study is just the first step at unleashing the enormous potential of generative AI in content generation. Future studies could also explore generating different types of items, such as conceptual problems.

Next, we examined the impact of giving students open access to isomorphic problem banks on students' exam performance. In both studies, there was no clear evidence that this approach considerably changed the difficulty of the problem on the exam, nor lead to widespread rote learning among students.

In both studies, the differences in CTT difficulty between the open bank problem and the transfer problem were less than 10%, which is much smaller compared to the differences between those problems and all other problems on the same exam, especially in Study 2. In addition, the 10% difference in CTT difficulty is comparable to the difference between two isomorphic versions of the same problem. Results from IRT analysis also confirmed those observations, indicating that the difficulty and discrimination of the open bank problem and transfer problem(s) are very similar compared to other problems on the same exam. In short, the results rejected our Hypothesis 1, which states that giving students open access to the problem bank would make the problems significantly easier on the exam.

Also on both exams, students' performance on open bank and transfer problems are correlated with each other, and the correlations are stronger than the average correlation between problems on the same exam. This goes against our second hypothesis that the performance will be uncorrelated due to students using rote memorization to answer the open bank problem alone. Furthermore, EFA analyses show that in both cases, the open bank problem and the transfer problem(s) load on the same factor. In particular, in Study 2 the open bank problem and the two transfer problems formed their own single factor, which also suggests that students are using similar processes to solve both the open bank problem and the transfer problem(s).

Finally, we observed mixed evidence for specificity in learning from problem banks (Hypothesis 3). In Study 1 and the near transfer case in Study 2 (between Q4 and Q6), the correlations are higher between mismatched versions of the open-bank and transfer problems, than between matched versions of the problems. Those observations did not support the hypothesis that students' learning from open problem banks are overly specific to the problems and have low ability to transfer to slightly different problem contexts. However, in the far transfer case in Study 2 (between Q6 and Q8), the matched versions are indeed more correlated than mismatched versions, which could indicate some level of specificity in student learning. However, it is also possible that the correlation between matched versions is the result of cognitive overload resulting from the complexity of Q8, or a priming effect from Q6.

Overall, the results of the study indicate that the open-bank approach's impact on the difficulty of the problems on an exam were quite small, as the tested problems remain some of the hardest problems on the exam. Given the significant potential benefits of the open bank approach, such as being able to widely share and infinitely reuse assessment problem banks without any concern of websites such as Chegg, it seems worthwhile to replace at least some of the items on traditional exams with open-isomorphic problem bank problems.

More importantly, if future assessment can consist entirely of open bank exam, students would be able to take the assessment at different times and have practically infinite number of attempts. Those assessments would enable authentic mastery-based learning, where students can take the assessment based on their level of content mastery instead of according to a pre-determined schedule.

**Caveats and Future Directions**

While the results of current study are encouraging for the use of open-isomorphic problem banks, there are a number of caveats and limitations that demand further investigated in future follow up studies.

First, the current study inferred the likelihood of rote-learning from students' performance on exam problems. Future follow-up studies could directly measure students' learning strategies by either survey, interview, or analyzing log data of students interacting with the open problem banks. In particular, different learning/practice strategies could be identified from log data utilizing machine learning methods such as those used in Zhang et. al. [58]. Those studies could provide direct evidence on students' learning strategy, and reveal how different strategies correlate with students' performance on the exam problems. Similarly, in this study we used the correlation of outcomes to infer whether students are using similar procedures to solve the two types of problems. Future studies could ask students to provide written solutions to the problems and examine whether there are qualitative differences on students' solutions for open-bank and transfer problems.

A second major caveat is that the results reported are limited to one problem type (numerical input) and 2-3 problems on each exam. With the advancement of AI assisted problem generation technique, it will be possible to generate larger numbers and a wider variety of problems. This will enable future studies to examine whether the current results will change if most or all of the problems on a given exam are open-bank problems, or when the problem banks contain one-step conceptual questions or simple calculation questions, instead of the multi-step numerical questions used in the current study. Furthermore, in the current study the problem banks are only available to students one week prior to the exam. Will the results change when the problems are available from the start of the semester? Research on those types of questions is needed to establish and continuously refine the open isomorphic problem bank approach as a reliable and efficient assessment method.

A third important unanswered question about the open isomorphic problem bank approach is whether all problems in the same bank can indeed be considered as isomorphic, or at least of very similar levels of difficulty, for every student. Results from this study show that the CTT difficulty of isomorphic problem variations can be different by about 10%. Is this amount of difference in difficulty acceptable, if students receive one problem that is randomly drawn from the problem bank on future exams? Can the impact of varying difficulty between isomorphic versions be reduced by non-conventional assessment practices, such as allowing multiple attempts, selecting more than one question from the bank, or allowing the test taker to select their own question? More importantly, can we design analysis methods to detect non-isomorphic items from student practice data, and remove or re-design those items before they are being implemented on an assessment? Finally, can we find better ways of defining and creating isomorphic problems that can strike a good balance between achieving relatively uniform difficulty and deterring rote-memorization or specific learning?

Finally, future studies need to pay special attention to whether the open isomorphic problem bank approach is, or can be made into, an equitable assessment approach for all students. In particular, it is possible that certain isomorphic variations may not be considered as isomorphic by some groups of students. For example, certain problem versions could involve context that are not familiar to students of certain cultural backgrounds (such as baseball or football references) or use vocabulary that maybe more challenging for non-native English speakers. While inherent biases are present in conventional assessments as well (for example, see [59]), we must strive to ensure that the AI assisted problem generation approach would serve to mitigate rather than exacerbate those existing inequalities. In this regard, the open problem bank approach could facilitate research on assessment equity, by allowing more data to be accumulated on problems in the problem bank, since the problem banks can be used by a much larger student population and for longer periods of time in higher stakes assessments such as exams. For example, future studies could develop methods to detect problems that are biased against certain demographic population by analyzing large amounts of students' practice data, and devise procedures to mitigate the impact of those biases. It may even be possible to determine a custom set of isomorphic assessment problems for each individual student, based both students' background information and their practice data.

In summary, even though there are still many unanswered questions, we believe that large open problem banks is a promising approach to harness the power of generative AI to transform existing practices of assessment in STEM disciplines.

**ACKNOWLEDGEMENT**

The authors would like to thank the Learning Systems and Technology team at UCF for developing the Obojobo online learning platform. This research is partly supported by NSF Grant No. DUE-1845436 and the College of Science Seed Funding Program at the University of Central Florida.